\documentclass[aps,reprint,twocolumn]{revtex4-2}
\usepackage[utf8]{inputenc}
\usepackage{amsmath,amssymb}
\usepackage{graphicx}
\usepackage{textgreek}

\usepackage{dcolumn}

\usepackage[utf8]{inputenc}
\usepackage[T1]{fontenc}
\usepackage{mathptmx}
\usepackage{etoolbox}
\usepackage{comment}
\usepackage{float}
\usepackage{hyperref}

\begin{document}
	
\title{Calibration of Microscope-coupled Fourier Transform Infrared Spectrometers for CW and Modulated Light Emission Measurements}

\author{Maxime Brazeau\textsuperscript{1}}

\author{Mathieu Giroux\textsuperscript{1}}

\author{Nada Boubrik\textsuperscript{1}}

\author{Raphael St-Gelais\textsuperscript{1,2}}
\email{raphael.stgelais@uottawa.ca}

\affiliation{\textsuperscript{1}Department of Mechanical Engineering, University of Ottawa}
\affiliation{\textsuperscript{2}Department of Physics, University of Ottawa}

\date{\today}

\begin{abstract}
	Measurement of low power infrared light emission spectra from microstructures can be challenging, but is of key importance in several research fields. Fourier transform infrared spectrometers (FTIR) can be used for characterizing such weak light emitters, but this requires additional custom user calibration compared to traditional FTIR measurements of, e.g., transmission or reflection. These calibration techniques are well documented for standalone FTIR instruments but not for microscope coupled-FTIRs, even though such an architecture greatly simplifies collection of light from micro and nano scale structures. We propose and demonstrate a calibration method for microsope-FTIRs based on the well-known emissivity of doped silicon at high temperature. With this method, we measure responsivity and noise floor of a recently installed microscope-FTIR instrument (Bruker\textsuperscript{\textcopyright} Invenio\textsuperscript{\textregistered} R coupled with a Hyperion II microscope), which is found to be within theoretically predicted values. The method is demonstrated for two different detectors (Mercury Cadmium Telluride and Indium Antimonide), in both continuous wave (CW) and modulated (step-scan) emission measurements mode. 
\end{abstract}

\maketitle

\section{\label{sec:intro}Introduction}

Measuring thermal emission spectra from micro and nano scale systems is particularly challenging due to their small size and inherently low levels of emitted radiation. Nevertheless accurate thermal emission measurements are critical for the development of micro-electro-mechanical systems (MEMS) in different fields such as gas detections \cite{Popa_2021, Lochbaum_2017}, chemical and biological sensors \cite{Pruessner_2025}, or for near field thermophotovoltaic (NFTPV) experiments \cite{kim_2015}.

With some instrument customization, it is possible to measure light emission from samples using a Fourier transform infrared (FTIR) spectrometer, but this feature is typically not calibrated by equipment manufacturers. FTIR spectrometers are instead more commonly used to measure the absorption, reflection and/or transmission of samples \cite{Griffiths_2007}. Calibration techniques for emission measurements have been developed and used in several previous work \cite{Xiao_2020, Xiao_2019, Dyachenko_2016, Wang_1997}, but they were not adapted to a microscope-FTIR system. Likewise, these calibration methods were not applied to modulated emission collected with a microscope-coupled-FTIR, commonly known as step-scan measurements. Such step-scan measurements allow measurement of a modulated emission signal only, thus removing the continuous wave (CW) thermal background \cite{Manning_2006, Andrews_2001}. 

Microscope-coupled FTIR assemblies are advantageous for collecting thermal light emission from on chip microstructures, but the microscope geometry complicates calibration. It requires a reference radiation source fitting under the objective, on the microscope platen, preventing the use of several calibrated light sources such laboratory blackbody radiators. 

Here we propose and demonstrate the calibration of a microscope-coupled FTIR using a heated piece of doped silicon, for which thermal emissivity models at high temperature are readily available. We rely on the model developed in \cite{Giroux_2024}, which accounts for optical absorption from dopants and thermally excited charge carriers in Silicon. 

\section{\label{sec:theory}Theory}
The spectral measured density measured by a typical FTIR is accessible to the user in units of arbitrary signal counts per wavenumber [counts/cm\textsuperscript{-1}]. We denote these measured spectra as $S_{\alpha,\mathrm{meas}}$ ($\tilde{\nu}$) [counts/cm\textsuperscript{-1}], with $\alpha$ denoting a given sample under test, and $\tilde{\nu}$ representing wavenumbers. Our first goal is to measure a responsivity function, denoted $m(\tilde{\nu})$ in units of counts/W. This function is such that dividing a measured spectrum by the response function results in a quantitative spectrum ($S_\alpha(\tilde{\nu})$, in W/cm\textsuperscript{-1}), in the same fashion as a typical optical spectrum analyser:
\begin{equation}
	S_{\alpha}(\tilde{\nu})=\frac{S_{\alpha, \mathrm{meas}}(\tilde{\nu})}{m(\tilde{\nu})} \mathrm{\left[\frac{W}{cm^{-1}}\right]}.
	\label{S_alpha}
\end{equation}

Here $S_{\alpha}(\tilde{\nu})$ is the spectral power density emitted by the sample under test $\alpha$ within the acceptance angle and area of the spectrometer, i.e., within its throughput \cite{Xiao_2019} $\theta$, in $\mathrm{m^2\cdot sr}$. 

To obtain the responsivity function $m(\tilde{\nu})$, we compare our measured signal $S_{Si,\mathrm{meas}}(\tilde{\nu})$ to a predictable intensity of radiation coupled inside the spectrometer, from a controlled thermal radiation emitter ($S_{\mathrm{Si,th}}$). This known emitter consists of a large area thermal emitter of known temperature $T$, and of known spectral emissivity $\epsilon(\tilde{\nu}, T)$. When such a source is placed under the microscope objective, the radiation coupled within the FTIR throughput $\theta$ is theoretically (th) predicted by \cite{Bergman_2020}:
\begin{equation}
	S_\mathrm{Si,th}(\tilde{\nu},T)=\theta\epsilon(\tilde{\nu}, T)B(\tilde{\nu}, T) \mathrm{\left[\frac{W}{cm^{-1}}\right]}.
	\label{S_th}
\end{equation}
In this case, $B(\tilde{\nu}, T)$ denotes Plank’s spectral radiance \cite{Griffiths_2007}:
\begin{equation}
	B(\tilde{\nu}, T)=\frac{2hc^2\tilde{\nu}^3}{e^{hc\tilde{\nu}/k_B T}-1} \mathrm{\left[\frac{W}{m^2 cm^{-1}sr}\right]}.
	\label{B_nu}
\end{equation}
where $h=6.626\times10^{-34}$ Js is the Planck constant, $k_B=1.381\times10^{-23}$ J/K is the Boltzmann constant and $c=2.998\times10^8$ m/s is the speed of light in vacuum. 

From this reference source, we obtain the responsivity $m(\tilde{\nu})$ by comparing the raw measured spectrum of the know surface $S_{\mathrm{Si,meas}}(\tilde{\nu})$ to its predicted value:
\begin{equation}
	m(\tilde{\nu})=\frac{S_\mathrm{Si,meas}(\tilde{\nu})}{S_\mathrm{Si,th}(\tilde{\nu})} \mathrm{\left[\frac{counts}{W}\right]}.
	\label{m_nu}
\end{equation}
In the common case where background radiation is non negligible, the differential procedure given in \cite{Xiao_2019} is used, allowing background cancellation by performing measurements at two radiator temperatures $T_1$ and $T_2$:
\begin{equation}
	m(\tilde{\nu})=\frac{S_\mathrm{Si,meas}(\tilde{\nu}, T_2)-S_\mathrm{Si,meas}(\tilde{\nu}, T_1)}{S_\mathrm{Si,th}(\tilde{\nu}, T_2)-S_\mathrm{Si,th}(\tilde{\nu}, T_1)}  \mathrm{\left[\frac{counts}{W}\right]}.
	\label{m_nu_T}
\end{equation}

Noteworthy, the final measured responsivity $m(\tilde{\nu})$ is an instrument parameter that should be independent of the chosen temperatures ($T_1$ and $T_2$), and of the throughput $\theta$. This will be confirmed experimentally in the next section.

The throughput of the FTIR (used in eq \ref{S_th}) is either limited by the diameter of the mirror and detectors, by the Jacquinot stop of the instrument, by the solid angle of the beam in the interferometer or, in a microscope-coupled FTIR, by the acceptance of the objective. In the present study, we reduce the aperture $D_\mathrm{ap}$ of the Jacquinot stop until we are aperture-limited, which yield a known throughput value:
\begin{equation}
	\theta=\frac{\pi D_\mathrm{ap}^2}{4}\times\Omega\; \mathrm{[m^2 sr]},
	\label{throughput}
\end{equation}
where $\Omega$ is the solid angle (in sr) between the aperture and its collection mirror. 

After having measured the responsivity $m(\tilde{\nu})$, we can measure the instrument noise spectral density ($S_\mathrm{N}$, in W/cm\textsuperscript{-1}) from a raw noise spectrum $S_{\mathrm{N,meas}}(\tilde{\nu})$:
\begin{equation}
	S_\mathrm{N}=\frac{S_\mathrm{N,meas}(\tilde{\nu})}{m(\tilde{\nu})} \mathrm{\left[\frac{W}{cm^{-1}}\right]}.
	\label{S_N}
\end{equation}

We can then compare measured noise ($S_\mathrm{N}$) to the predicted noise floor ($N$) obtained from the detector specific detectivity ($D^*$) and from the ambient thermal background. The noise floor obtained from the detector specific detectivity is given by \cite{Griffiths_2007}:
\begin{equation}
	N_\mathrm{det}=\frac{\sqrt{A_\mathrm{D} f_\mathrm{D}}}{D^*\Delta\tilde{\nu}} \mathrm{\left[\frac{W}{cm^{-1}}\right]}.
	\label{N_det}
\end{equation}
In this case $f_\mathrm{D}$ is the detector electronic sampling bandwidth. We assume this value to be approximately equal to the scanner velocity (e.g., 20 kHz) when the FTIR is used in continuous scanning mode, and to the lock-in amplifier demodulation bandwidth when performing step-scan measurements. $D^*$ is the specific detectivity given by the manufacturer \cite{Bruker_2019}, and $A_\mathrm{D}$ is the detector area.

In continuous scanning mode, the inside of the FTIR acts like a blackbody source at room temperature, which contributes to the instrument background according to:
\begin{equation}
	N_\mathrm{bgnd}=\theta_\mathrm{n} B(\tilde{\nu},T_\mathrm{room}) \mathrm{\left[\frac{W}{cm^{-1}}\right]},
	\label{N_bgnd}
\end{equation}
where $\theta_\mathrm{n}$ is the throughput of the background. As discussed in Appendix \ref{appendix_A} it can differ from the throughput of the FTIR and should be calculated separately. We can finally combine eq \ref{N_det} and \ref{N_bgnd} in a total noise floor using
\begin{equation}
	N_\mathrm{tot}^2=N_\mathrm{det}^2+N_\mathrm{bgnd}^2.
	\label{N_tot}
\end{equation}

In short, we can predict the instrument noise floor using eq \ref{N_det}-\ref{N_tot} and compare it with the measured noise floor of eq \ref{S_N}. To do so, we must however first measure the instrument responsivity $m(\tilde{\nu})$, using the procedure described in the next section.

\section{\label{sec:methods}Methods}

\begin{figure}
	\includegraphics{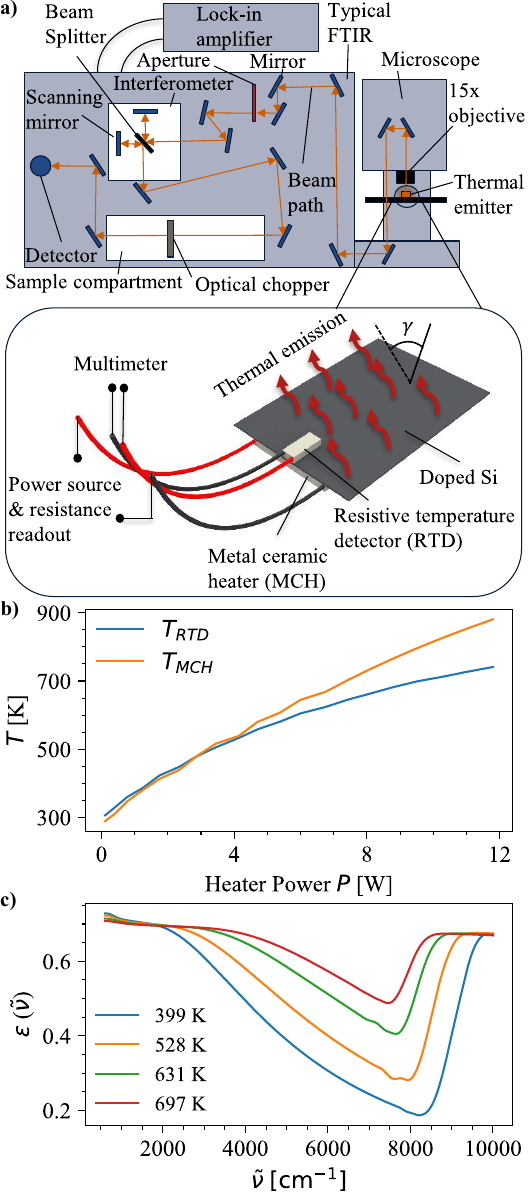}
	\caption{\label{fig:schem} a) Schematic of the microscope-coupled-FTIR, reproduced from \cite{bruker2015invenioR}, with a zoomed in view of the controlled thermal emitter. b) Temperature of the thermal emitter measured from the resistance change in the metal ceramic heater and from a resistance temperature detector. c) Calculated spectral emissivity of the emitter at various temperatures.}
\end{figure}

We characterize a Bruker\textsuperscript{\textcopyright} Invenio\textsuperscript{\textregistered} R FTIR paired with a Hyperion II microscope, installed new in August 2023. The microscope is primarily designed for measuring the reflectivity and/or transmissivity of samples. A factory modification allows using the sample under the microscope as a light source, paired with the FTIR internal (interchangeable) detectors. Fig.~\ref{fig:schem}a illustrates this beam path in a simplified manner. Thermal emission from the sample is collected by a 15$\times$ magnification Schwarzschild objective \cite{bruker2020hyperion} and then passed through the aperture stop, where its amplitude can be adjusted by varying the aperture size. To determine the limiting aperture $D_\mathrm{ap}$ used for calculating the throughput in eq \ref{throughput}, we reduce the aperture size until the signal amplitude reduces (i.e., until the throughput $\theta$ is limited by the aperture), see Appendix \ref{appendix_A}. 

After passing the aperture, the beam enters the scanning Michelson interferometer equipped with either a KBr or a CaF\textsubscript{2} beam splitter. A KBr beam splitter is used for mid-IR measurements (350 --- 8000 cm\textsuperscript{-1}) and CaF\textsubscript{2} is used for near-IR measurements (1200 --- 15500 cm\textsuperscript{-1}) \cite{bruker2015invenioR}. Finally, the signal is directed to the detector. We use two different detectors, a mercury cadmium tellurium (MCT) in the range of $\tilde{\nu}=600$ cm\textsuperscript{-1} to $\tilde{\nu}=12 000$ cm\textsuperscript{-1} and a InSb detector for $\tilde{\nu}=1850$ cm\textsuperscript{-1} to $\tilde{\nu}=10 000$ cm\textsuperscript{-1} (part number D316/B and D413/B, respectively). Their peak nominal detectivities are respectively $D^*>2\times10^{10}$ cm Hz\textsuperscript{1/2} W\textsuperscript{-1} at 800 cm\textsuperscript{-1} and $D^*>1.5\times10^{11}$ cm Hz\textsuperscript{1/2} W\textsuperscript{-1} at 2200 cm\textsuperscript{-1}. \cite{ Bruker_2019} 

Fig.~\ref{fig:schem}a illustrates our calibrating radiator, which consists of a piece of Si bonded to a metal-ceramic heater (MCH). The MCH is connected to a power supply (Rigol DB831) to control the applied voltage, and hence the temperature. The Si is N-doped with arsenide and has a resistance ranging from 0.001 to 0.005 $\mathrm{\Omega\cdot cm}$. We use silver paste (PELCO\textsuperscript{\textregistered} 16047) to bond the Si to the MCH, which enables low heat transfer resistance and high temperature stability. 

During heating, we measure the temperature of the Si radiator using two methods. First, a resistance temperature detector (RTD) (Innovative Sensor Technology 2952-P1K0.520.6W.A.010-ND) \cite{RTD} is bonded to the surface of the Si and connected to a multimeter (Keithley DMM 6500). The second method consists of measuring the change of electrical resistance of the MCH itself during heating. In this case we calculate the change in temperature using \cite{Giroux_thesis}:
\begin{equation}
	\Delta T = \frac{1}{\alpha_{\mathrm{th}}}\frac{\Delta R}{R_0}
\end{equation}
where $\Delta R$ is the change of resistance, $R_0$ is the resistance at room temperature, $\Delta T$ is the change in temperature and $\alpha_\mathrm{th}=0.0045$/$^{\circ}$C \cite{Giroux_thesis} is the temperature coefficient of resistance of the MCH. $\alpha_\mathrm{th}$ is measured in a separate experiment using a hot plate to set $\Delta T$, and then measuring the change in resistance, as described in \cite{Giroux_thesis}. The temperature measured by the two techniques initially follows similar trends, but deviates at high temperatures, as shown in Fig.~\ref{fig:schem}b. This difference is expected at high heating power, whereas radiative heat transfer becomes significant, increasing the temperature difference between the heater and the radiator surface. Since the radiative surface is what matters in eqs \ref{S_th}-\ref{m_nu_T}, we primarily rely on $T_\mathrm{RTD}$ for our calculations.

The permittivity of Si is predicted from its dielectric function, calculated using the model proposed in \cite{Giroux_2024}. The model incorporates the lattice and interband contributions from \cite{Lee_2005}, where interband absorption is approximated using the semi-empirical discrete energy-level model described in \cite{Timans_1993}. The free carrier contribution is represented by the Drude model from \cite{Fu_2006}, with mobility, decay rates, and donor and acceptor contributions taken from \cite{Basu_2009}. Moreover, the model considers optical absorption from dopants as well as from thermally excited charge carriers by incorporating the temperature dependence of decay rates and carrier concentration. 

With this permittivity model, we calculate (Fig.~\ref{fig:schem}c) the silicon spectral directional emissivity $\epsilon(\tilde{\nu},\gamma)$ function at an angle $\gamma$. The spectral directional emissivity is a unitless number representing the ratio of emited energy by a surface compared to a perfect black body emitter. Using Kirchhoff's law of radiation it can be related to the absorptivity and reflectivity of a surface \cite{Bergman_2020}:
\begin{equation}
	\epsilon(\tilde{\nu},\gamma)=\alpha(\tilde{\nu},\gamma)=1-R(\tilde{\nu},\gamma),
\end{equation}
where $\alpha(\tilde{\nu},\gamma)$ is the absorption coefficient of the heated silicon piece, and $R(\tilde{\nu},\gamma)$ is the reflection coefficient of the exposed silicon surface. This reflection coefficient $R(\tilde{\nu},\gamma)$ is calculated using an optical multilayer calculation algorithm \cite{Edalatpour_2013}, considering air as the incident medium followed by a 280 \textmu m thick silicon layer, and terminated by semi-infinite silver. The permittivity of silver is taken from \cite{Yang_2015}. This semi infinite last layer accounts for silver paste at the heater/silicon interface, which is not expected to allow any light transmission, hence the semi-infinite assumption. Once the directional emissivity $\epsilon(\tilde{\nu},\gamma)$ is calculated, we can obtain the total spectral emissivity $\epsilon(\tilde{\nu})$  (needed in eq \ref{S_th}) by integrating $\epsilon(\tilde{\nu},\gamma)$ over all the emission angles within the FTIR throughput. We note however (see Appendix \ref{appendix_B}) that this integral is very weakly dependent on the integration range of $\gamma$, such that we could also simply use the diffuse surface approximation, i.e., $\epsilon(\tilde{\nu})\approx\epsilon(\tilde{\nu},0)$, where $\epsilon(\tilde{\nu},0)$ denotes the spectral directional emissivity at normal incidence.

For step-scan calibration measurements, we use an optical chopper, placed in the sample compartment, to modulate the CW light emitted by the silicon emitter (see Fig.~\ref{fig:schem}a). In this case, the AC output port of the detector is connected to a lock-in amplifier (Zurich Instrument MFLI 500 kHz/5 MHz, 60 MSa/s) which demodulates and amplify the signal at the chopper frequency, before returning it to the FTIR built-in digitizer. Returning only the demodulated signal to the FTIR allows suppression of unmodulated thermal background \cite{Manning_2006}, which is useful for measuring weak modulated light sources. We have tested modulation frequencies in the 0 Hz to 1 kHz range and found that a chopping frequency of 1 kHz allows for better suppression of background and low frequency noise. We also find that amplifying the signal by a factor of 100 at the output of the lock-in amplifier reduces noise by making the FTIR digitizer noise negligible. Other parameters of the step-scan measurements are the lock-in amplifier’s time constant of 32 ms, a settling time of 150 ms after FTIR mirror displacement followed by, an acquisition time of 150 ms, and a resolution of $\Delta\tilde{\nu}=32$ cm\textsuperscript{-1}.

It should be noted that locating the chopper in the sample compartment (as in Fig.~\ref{fig:schem}a) modulates part of the unwanted thermal background, in addition to light emitted by the heated silicon surface. It is therefore essential to use the background cancelling method (see eq \ref{m_nu_T}) in this case. 

To be throughput limited we need an aperture size $D_\mathrm{ap}$ of 4 mm in CW and 1.5 mm in step-scan for the MCT detector (see Appendix \ref{appendix_A}). There is a difference in throughput between the CW measurement and the step-scan due to the use of a perforated plate (4.5 mm diameter) in the sample compartment during the step-scan measurements. This plate is used to reduce the beam diameter such that it fits within the optical chopper blades placed in the sample compartment (see Fig.~\ref{fig:schem}a). These 4 mm and 1.5 mm aperture correspond to throughputs of $7.24\times10^{-7}$ $\mathrm{m^2\cdot sr}$ and $1.01\times10^{-7}$ $\mathrm{m^2\cdot sr}$, respectively, considering that the beam is collimated to a beam diameter $D_{beam}=30$ mm by a mirror of a $f=110$ mm focal length after the aperture.

Following a similar approach for the InSb detector, we need an aperture of 3 mm and 2.5 mm in CW and in step-scan respectively to be aperture limited (see Appendix \ref{appendix_A}). Leading to a throughput of $4.07\times10^{-7}$ $\mathrm{m^2\cdot sr}$ and $2.83\times10^{-7}$ $\mathrm{m^2\cdot sr}$ in CW and step-scan, respectively.

To summarize, the calibration procedure consist of six main steps:
\begin{enumerate}
	\item Choose the proper detector/beam splitter combinaision for NIR or MIR measurements. (e.g., MCT and KBr for MIR or InSb and CaF\textsubscript{2} for NIR)
	\item Determine the limiting aperture by gradually reducing the aperture size until the measured signal amplitude of the heated control emitter becomes aperture-limited. Then, calculate the instrument throughput using Eq. \ref{throughput}.
	\item Measure the raw thermal emission spectrum of the control radiation emitter ($S_\mathrm{Si, meas}(\tilde{\nu})$) at two different temperatures ($T_1$ and $T_2$).
	\begin{enumerate}
		\item For step-scan calibration, use a lock-in amplifier and optical chopper wired as in Fig.~\ref{fig:schem}a.
	\end{enumerate}
	\item Calculate the response function $m(\tilde{\nu})$ using eq \ref{m_nu_T}.
	\item Measure the raw emission spectrum of a heated sample of interest ($S_{\alpha, \mathrm{meas}}(\tilde{\nu})$) placed under the microscope. For weak light emission measurement, the limiting aperture may be increased relative to the one used in step 2.
	\item Using eq \ref{S_alpha} and the response function $m(\tilde{\nu})$ obtained in step 4, calculate the calibrated sample emission spectrum ($S_\alpha(\tilde{\nu})$). The obtained spectrum (in W/cm\textsuperscript{-1}) quantifies light emitted by the sample within the instrument throughput $\theta$ set in step 5.
\end{enumerate}

\section{\label{sec:results}Results}

\begin{figure}
	\includegraphics{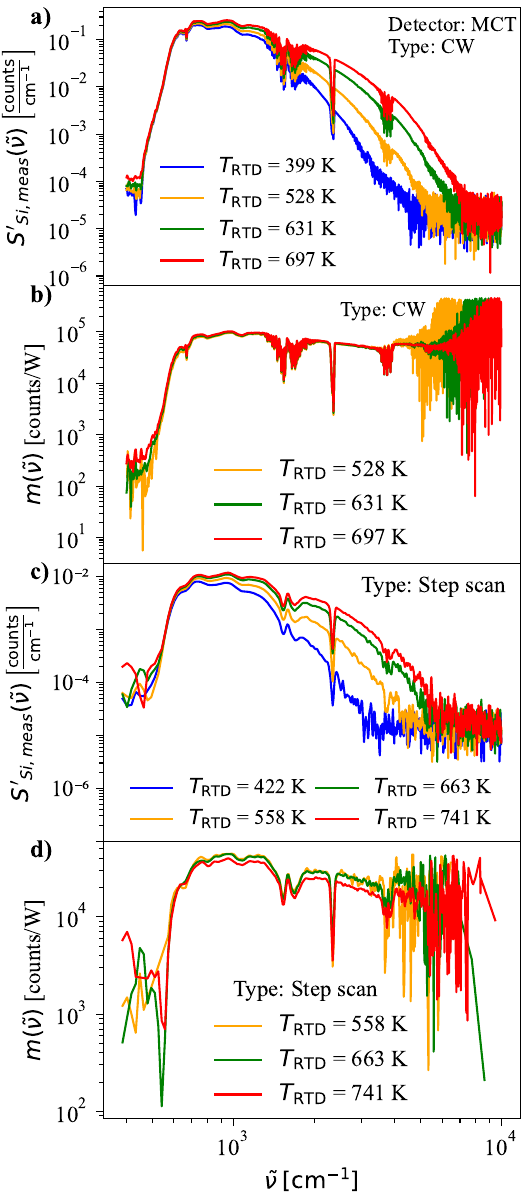}
	\caption{\label{fig:MCT} a) Measured raw emission signal $S_{\mathrm{Si,meas}}$ collected with the MCT detector in CW mode. b) Response function $m(\tilde{\nu})$ of the MCT detector in CW mode. c) Measured raw emission signal $S_{\mathrm{Si,meas}}$ collected with the MCT detector in step-scan mode. d) Response function $m(\tilde{\nu})$ of the MCT detector in step-scan mode. All measurements are taken at various emitter temperatures ($T_\mathrm{RTD}$).}
\end{figure}

\begin{figure}
	\includegraphics{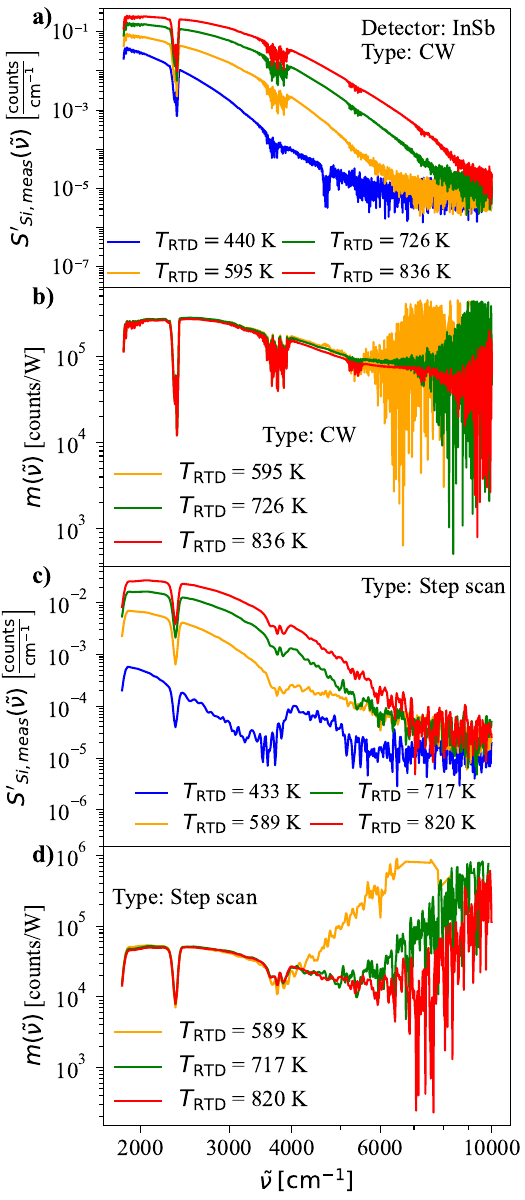}
	\caption{\label{fig:InSb} a) Measured raw emission signal $S_{\mathrm{Si,meas}}$ collected with the InSb detector in CW mode. b) Response function $m(\tilde{\nu})$ of the InSb detector in CW mode. c) Measured raw emission signal $S_{\mathrm{Si,meas}}$ collected with the InSb detector in step-scan mode. d) Response function $m(\tilde{\nu})$ of the InSb detector in step-scan mode. All measurements are taken at various emitter temperatures ($T_\mathrm{RTD}$).}
\end{figure}

In Fig.~\ref{fig:MCT} and Fig.~\ref{fig:InSb}, we measure the response function $m(\tilde{\nu})$ of the instrument when equipped with two different detectors (MCT and InSb, respectively), in CW and in step-scan. To do so, we first measure the raw emission spectra ($S_{\mathrm{Si,meas}}$) of the silicon radiator at four different temperatures. We then proceed to background cancellation according to eq \ref{m_nu_T}. In this case, $T_1$ is the lowest measured emission temperature (e.g., $T_\mathrm{RTD}=399$ K in Fig.~\ref{fig:MCT}a) and $T_2$ is successively set to the other three $T_\mathrm{RTD}$ emission temperatures. 

For the MCT detector, Fig.~\ref{fig:MCT}a and Fig.~\ref{fig:MCT}c show the raw measured signal $S_{\mathrm{Si,meas}}(\tilde{\nu})$ of the silicon emitter at various RTD temperatures ($T_\mathrm{RTD}$) for the CW and step-scan measurements respectively. Peaks around 1400 to 1800 cm\textsuperscript{-1} and 3500 to 4000 cm\textsuperscript{-1} are caused by water absorption while peaks around 2200 to 2400 cm\textsuperscript{-1} are caused by CO\textsubscript{2} absorption \cite{Smal_2014}. As $T_{RTD}$ increases, the device emits more light at higher wavenumber. This is expected and predicted in the theoretical value $S_{\mathrm{Si,th}}$ (see Appendix \ref{appendix_B}). Using the curves from Fig.~\ref{fig:MCT}a and c, with eq. \ref{m_nu_T}, we calculate the response function $m(\tilde{\nu})$ at different temperatures. In this case, we use $T_1=T_\mathrm{RTD}=440$ K in CW and $T_1=T_{RTD}=433$ K in step-scan. We then use $T_2=528$ K, 631K and 697 K in CW, and $T_2=558$ K, 663K and 741 K in step-scan.

Still for the MCT detector, Fig.~\ref{fig:MCT}b and Fig.~\ref{fig:MCT}d show the response function $m(\tilde{\nu})$ with the various emitter temperatures in CW and step-scan, respectively. Importantly, the response functions are essentially temperature independant, as expected from eq 5. The lowest noise, highest temperature curve (e.g., at 697 K in Fig.~\ref{fig:MCT}b) can be used for calibrating the emission of any sample, regardless of its temperature. The lower temperature curves are included in the figures to validate the method: i.e., a dependence in temperature would indicate a problem with our calibration. We note a maximum around 800 cm\textsuperscript{-1} which is consistent with detectivity charts for the MCT detector \cite{ Bruker_2019}. Importantly, both response functions are essentially temperature-independent, as expected in eq \ref{m_nu_T}. This is an important confirmation of the validity of our method since errors in, for example, the measurement of the radiator temperature would cause a mismatch of the different response functions. At higher $T_\mathrm{RTD}$, there is less noise at higher wavenumbers for the response function $m(\tilde{\nu})$. This is expected from the higher emission in the NIR range as shown in Fig.~\ref{fig:MCT}a and c.

Following a similar process for the InSb detector, Fig.~\ref{fig:InSb}a and Fig.~\ref{fig:InSb}c show the measured signal of the calibrated emitter at various emitter temperatures in CW and step-scan measurements respectively. In CW we use $T_1=T_\mathrm{RTD}=440$ K and $T_2=726$ K, 595 K and 836 K. In step-scan $T_2=589$ K, 717 K and 820 K. 

Conversely, Fig.~\ref{fig:InSb}b and Fig.~\ref{fig:InSb}d shows the response function $m(\tilde{\nu})$ in CW and step-scan, respectively, for the InSb detector. Again, the peak around 2200 cm\textsuperscript{-1} matches detectivity charts for InSb detectors \cite{ Bruker_2019}. In Fig.~\ref{fig:InSb}d the $T_{RTD}=589$ K curve diverges from the other response at $\tilde{\nu}>4000$ cm\textsuperscript{-1} due to insufficient optical signal for calibration. Otherwise, response function behaves similarly as for the MCT detector: amplitudes are constant at various temperatures, and noise at high wavenumber decreases as the temperature increases. 

\begin{figure}
	\includegraphics{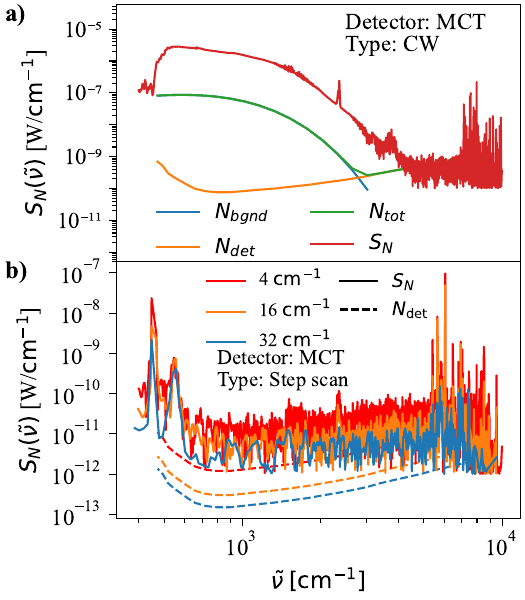}
	\caption{\label{fig:noise_MCT} a) Theoretical and measured noise of the FTIR equipped with the MCT detector, in CW mode, including the detector noise and the thermal background. The resolution is $\Delta\tilde{\nu}=4$ cm\textsuperscript{-1} b) Theoretical and measured noise of the MCT detector, in step-scan for multiple resolutions. In step-scan we only include the detector noise since there is no thermal background.}
\end{figure}

\begin{figure}
	\includegraphics{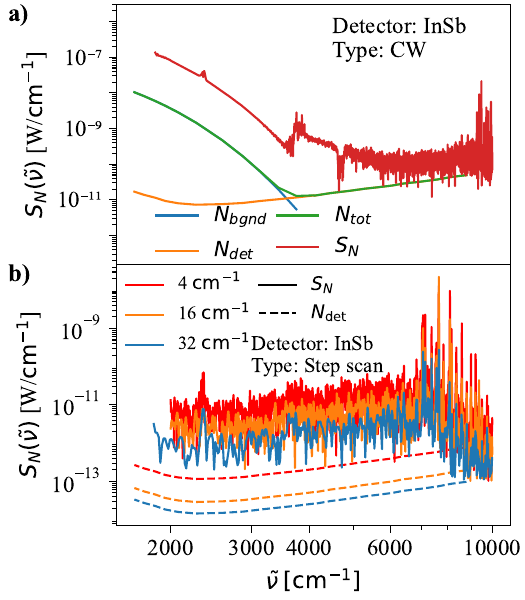}
	\caption{\label{fig:noise_InSb} a) Theoretical and measured noise of the FTIR equipped with the InSb detector, in CW mode, including the detector noise and the thermal background. The resolution is $\Delta\tilde{\nu}=4$ cm\textsuperscript{-1} b) Theoretical and measured noise of the InSb detector, in step-scan for multiple resolutions. In step-scan we only include the detector noise since there is no thermal background.}
\end{figure}

Importantly, it should be noted that the instrument response function shown in Fig.~\ref{fig:MCT} - \ref{fig:InSb} are throughput-independent due to the normalization performed in eq \ref{S_alpha} - \ref{m_nu}. In other words, $m(\tilde{\nu})$ is the response function to light coupled within the instrument throughput, whatever this throughput may be. In Appendix \ref{appendix_A}, we confirm that the response function is independent of the throughput, provided that we know what is limiting throughput.

We can then combine the response function for both detectors with the instrument self-noise to quantify the instrument noise floor $S_\mathrm{N}$ (see eq \ref{S_N}). We take a background measurement of the unheated sample at room temperature $S_{\mathrm{N,meas}}$ ($T_\mathrm{room}\approx300$ K) and calibrate this signal using the response function $m(\tilde{\nu})$ with eq \ref{S_alpha}. The noise floor are shown in Fig.~\ref{fig:noise_MCT} – \ref{fig:noise_InSb}a in CW, for respectively the MCT and InSb detectors, with a resolution $\Delta\tilde{\nu}=4$ cm\textsuperscript{-1}. Likewise, noise floor measurements in step-scan are presented in Fig.~\ref{fig:noise_MCT} – \ref{fig:noise_InSb} b for the same two detectors with a resolution $\Delta\tilde{\nu}=32$ cm\textsuperscript{-1}. Note that noise floors for other resolutions can be inferred from the data in Fig.~\ref{fig:noise_MCT} - \ref{fig:noise_InSb} using the relations given in eq. \ref{S_N} - \ref{N_tot}. In particular, noise floors spectra limited by detector noise (e.g., step-scan) can be expected to scale as $\Delta\tilde{\nu}^{-1}$ (eq \ref{N_det}) while noise floors are background limited (i.e., in CW) are resolution-independent (eq \ref{N_bgnd}). 

The measured noise floor $S_\mathrm{N}$ are consistent with theoretical predictions. The theoretical noise $N_\mathrm{tot}$ comprises the thermal room temperature background $N_\mathrm{bgnd}$ (in CW only) and the detector noise $N_\mathrm{det}$ (for both CW and step-scan). In CW, at lower wavenumbers ($600$  $\mathrm{cm^{-1}} < \tilde{\nu} < 4500$ $\mathrm{cm^{-1}}$) noise is limited by the thermal background $N_\mathrm{bgnd}$, while at  higher wavenumber $\tilde{\nu}>4500$ $\mathrm{cm^{-1}}$) it is limited by the detectors noise $N_\mathrm{det}$. In step-scan, we do not have $N_\mathrm{bgnd}$ due to the nature of the measurement, hence we are limited by $N_\mathrm{det}$ for the entire spectra. Note that detector noise $N_\mathrm{det}$, are nominal values taken from \cite{Bruker_2019} (see section \ref{sec:methods} for the responsitivity of detectors) and may vary for one detector to another. Theoretical detectors noise in Fig.~\ref{fig:noise_MCT} - \ref{fig:noise_InSb} should therefore be treated as an estimate and is not meant for precise quantitative comparisons. 

For CW measurements Fig.~\ref{fig:noise_MCT} – \ref{fig:noise_InSb} a, the measured noise is approximately a factor $\approx20$ above the predicted noise $N_\mathrm{tot}$ for the MCT detector and $\approx12$ for the InSb detector. This can likely be explained by optical losses in the instrument. As stated in \cite{Griffiths_2007}, a typical standalone FTIR has a light collection efficiency of 10\%, while additional losses can reasonably be expected within the microscope attachment.  

In step-scan (Fig.~\ref{fig:noise_MCT} – \ref{fig:noise_InSb} b), the theoretical noise is only composed of the detector noise $N_\mathrm{det}$ since the unmodulated background is rejected by lock-in detection. These measurement are 1-2 orders of magnitude above the nominal detector noises, which is consistent with the collection efficiency observed in CW.

We note that noise levels are significantly lower in step-scan than in CW. This difference is partly due to the absence of thermal background in the step-scan measurements at lower wavenumbers. However, at higher wavenumbers, the discrepancy arises from differences in averaging time and measurement frequency. The CW noise is measured with a resolution of 4 cm\textsuperscript{–1}, a mirror frequency of $f_D=20$ kHz, and averaging over 32 scans for a total measurement time of 20 s. In step-scan mode, various resolutions are used, with a lock-in time constant of 32 ms or $f_D=4.97$ Hz, a mirror stabilization time of 150 ms, and each interferogram data point is averaged for 150 ms. For the same resolution as in CW ($\Delta\tilde{\nu}=4$ cm\textsuperscript{-1}), our measurement time amounts to 1000 s in step-scan, i.e., 50 times longer than in CW. This longer sampling time explains the lower noise floor for step-scan measurements, as expected with noise scaling with $f_D^{1/2}$ in eq \ref{N_det}.

As a final sanity check of our calibration method, we measure light emission from light emitting diodes (LED) placed directly under the microscope. For CW measurements, we use an LED emitting at 6060 cm\textsuperscript{-1} (Marktech Optoelectronics MTE5116N5) with a nominal emission power of 5 mW \cite{datasheet_LED}. This near-infrared wavelength is chosen for CW measurement to ensure emission above the thermal background. Conversly, for step-scan measurements where the background is eliminated, we use an LED emitting at 2325 cm\textsuperscript{-1} (Hamamatsu L15895-0430MA) with a nominal emission power of 0.8 mW \cite{LED_4_3}. Since the emission power and cone angle of the LEDs are not calibrated separately, we cannot draw quantitative conclusions from these measurements and we therefore present then as supplementary material in Appendix \ref{appendix_C}. The data nevertheless indicates realistic emitted power values below the LED nominal emission value, as expected from not matching their emission cone with the FTIR throughput. 

\section{\label{sec:conclusion}Conclusion}
We demonstrated a simple practical method for calibrating a microscope-coupled FTIR for light emission measurements in both CW and step-scan modes. The response function $m(\tilde{\nu})$ also enables us to determine the instrument noise floor in emission mode, for near and mid-infrared detectors. These noise floors are within theoretically predicted values considering typical FTIR collection efficiencies. 

\begin{acknowledgments}
	The authors would like to thank Sergey Shilov from Bruker\textsuperscript{\textcopyright} for his insightful feedback and support during the development of this calibration method.
\end{acknowledgments}

\bibliography{Refs}

\clearpage

\appendix

\setcounter{section}{0} 
\renewcommand{\thesection}{\Alph{section}}

\section{Throughput Measurement}\label{appendix_A}

\renewcommand{\thefigure}{\thesection\arabic{figure}}
\renewcommand{\theequation}{\thesection\arabic{equation}}
\setcounter{figure}{0}
\setcounter{equation}{0}

The throughput of the FTIR is limited by either the diameter of the mirrors/detector, by the Jacquinot stop (i.e., aperture), by the acceptance angle of the microscope objective for microscope-coupled FTIR, or by the solid angle of the beam in the interferometer. The throughput of the detector (or aperture) is calculated as \cite{Griffiths_2007}:
\begin{equation}
	\theta=\Omega A=2\pi\left(1-\cos\alpha_\mathrm{M}\right)A,
	\label{theta_D}
\end{equation}
where A is the area of the detector (or aperture), $\Omega$ is the solid angle, and $\alpha_\mathrm{M}$ is the maximum half angle of the focused beam. 

The throughput of the solid angle of the beam in the interferometer can be calculated as \cite{Griffiths_2007}:
\begin{equation}
	\theta_\mathrm{I}=\Omega_\mathrm{I}A_\mathrm{M}=\frac{2\pi \Delta\tilde{\nu}A_\mathrm{M}}{\tilde{\nu}_\mathrm{max}},
	\label{theta_I}
\end{equation}
where $\Omega_\mathrm{I}$ is the solid angle of the beam of the interferometer, $A_\mathrm{M}$ is the area of the mirror in the interferometer, $\tilde{\nu}_\mathrm{max}$ is the maximum wavenumber measured and $\Delta\tilde{\nu}$ is the resolution. 

It is difficult to predict exactly what is limiting the throughput. For example, we do not know the throughput of the microscope objective. Likewise, in step-scan, we use a perforated plate to limit the beam diameter to within the chopper wheels, which will also add another variable to the throughput. Therefore, we impose the limiting throughput by reducing the aperture size until the signal becomes aperture-limited. To calculate the throughput, we then use eq \ref{theta_D} with $A_D$ equal to the aperture area and the maximum half angle is given by:
\begin{equation}
	\tan\alpha_\mathrm{M}=\frac{D_\mathrm{beam}}{2f},
	\label{alpha}
\end{equation}
where $D_\mathrm{beam}\approx30$ mm is the diameter of the beam following the aperture and $f$ is the focal length of the collection mirror following the aperture. 

\begin{figure}
	\includegraphics{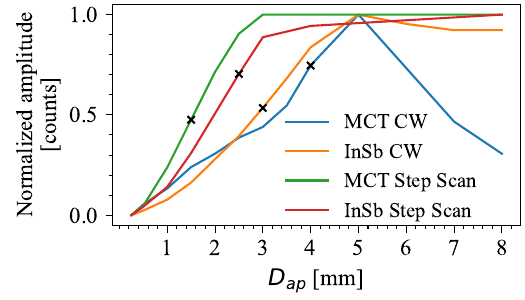}
	\caption{\label{fig:apperture} Normalized amplitude in function of aperture size for the MCT and InSb detectors in CW and in step-scan. The response function is calculated using an aperture of 4 mm for the MCT and 3 mm for the InSb detectors in CW. In step-scan we used an aperture of 1.5 mm and 2.5 mm for the MCT and InSb detectors respectively.}
\end{figure}

To assess the limiting aperture, for both CW and step-scan measurements, we place the controlled emitter at a high temperature (i.e., 400 K) under the microscope. Then we measure the amplitude of the signal starting with the largest aperture, gradually reducing it to the smallest aperture available in our instrument (i.e., from $D_\mathrm{ap}=8$ mm to $D_\mathrm{ap}=0.25$ mm). Fig.~\ref{fig:apperture} shows the normalized signal amplitude as function of the aperture size, where "$x$" symbol denotes the aperture used for $m(\tilde{\nu})$ measurements. 

For the CW measurements, the limiting aperture for the MCT detector is \textasciitilde5 mm, thus we use a 4 mm aperture for our measurements to ensure we are aperture limited. Similarly, for the InSb detector, the limiting aperture is also \textasciitilde5 mm and we use a 3 mm aperture. In step-scan, the limiting aperture for the MCT detector is \textasciitilde3 mm, and we use a 1.5 mm aperture. For the InSb detector, the limiting aperture is \textasciitilde4 mm, and we use a 2.5 mm aperture. Note that, as discussed in the main text, the final chosen aperture is not critically important as long as the system is aperture-limited.

As expected, the limiting aperture in step-scan is different from the CW measurements. In step-scan, we use a chopper in the sample compartment to modulate the signal (see Fig.~\ref{fig:schem}a). At this location, the beam diameter is larger than the chopper blades and we are unable to properly modulate the signal. Hence, we place a perforated plate in front of the chopper to reduce the diameter of the beam, reducing the signal amplitude and changing the limiting aperture.

When we are aperture limited, we calculate the throughput using eq \ref{theta_D} with $A=\pi D_\mathrm{ap}^2/4$. The maximum half angle is then calculated with eq \ref{alpha} using $D_\mathrm{beam}=30$ mm and a focal length of $f=110$ mm, resulting in a maximum half angle of $\alpha_\mathrm{M}\approx7.86^{\circ}$. 

This results in a throughput of $7.24\times10^{-7}$ $\mathrm{m^2\cdot sr}$ with $D_\mathrm{ap}=4$ mm and $1.02\times10^{-7}$ $\mathrm{m^2\cdot sr}$ with $D_\mathrm{ap}=1.5$ mm for the MCT in CW and step-scan mode respectively. For the InSb detector we have a throughput of $4.07\times10^{-7}$ $\mathrm{m^2\cdot sr}$ with $D_\mathrm{ap}=3$ mm and $2.82\times10^{-7}$$\mathrm{m^2\cdot sr}$ with $D_\mathrm{ap}=2.5$ mm in CW and step-scan mode respectively.

\begin{figure}
	\includegraphics{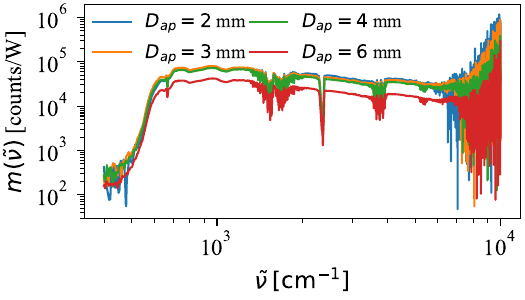}
	\caption{\label{fig:apperture_test} Response function $m(\tilde{\nu})$ for various aperture size for the MCT detector in CW mode. When aperture-limited (i.e., for 2, 3 and 4 mm), the response functions are superimposed, as expected.}
\end{figure}

Fig.~\ref{fig:apperture_test} shows the response function $m(\tilde{\nu})$ for various different aperture size (2, 3, 4 and 6 mm) for the MCT detector in CW mode. For apertures of 2, 3, and 4 mm, we are aperture limited, while for the 6 mm aperture, we are limited by something else in the instrument. Fig.~\ref{fig:apperture_test} therefore confirms that the response function $m(\tilde{\nu})$ is independent of throughput (2, 3, and 4 mm) as long as the throughput is known and accounted for in eq \ref{S_th}.

In eq \ref{S_N}, background noise has different throughput $\theta_\mathrm{n}$ since this noise originates from modulation of the background light by the interferometer, which is independent of the aperture stop. In this case, we calculate $\theta_n$ using beam size of $D_\mathrm{beam}\approx30$ mm and the focusing mirror before the detector of focal length of $f=33$ mm, leading to a throughput of $\theta_n=5.63\times10^{-7}$ $\mathrm{m^2\cdot sr}$.

\setcounter{section}{1} 
\renewcommand{\thesection}{\Alph{section}}

\section{Theoretical Emission of the Radiator}\label{appendix_B}

\renewcommand{\thefigure}{\thesection\arabic{figure}}
\renewcommand{\theequation}{\thesection\arabic{equation}}
\setcounter{figure}{0}
\setcounter{equation}{0}

In this section we derive the theoretical spectral emission $S_\mathrm{th}(\tilde{\nu})$ (see eq \ref{S_th}) of the controlled emitter used to calculate the response function $m(\tilde{\nu})$. 

Firstly, we show that the emissivity is weakly dependent on the angle $\gamma$, allowing us to use the diffused surface approximation $\epsilon(\tilde{\nu}, \gamma)=\epsilon(\tilde{\nu}, 0)$. To do so, the spectral hemispherical emissivity of the multi-layer (i.e., Air/Si/Ag) is calculated as \cite{Bergman_2020}:
\begin{equation} 
	\epsilon(\tilde{\nu},T)=\frac{\int^{\phi_\mathrm{max}}_{0}\int^{\gamma_\mathrm{max}}_{0}\epsilon(\tilde{\nu}, \gamma, \phi, T)\cos{\gamma}\sin{\gamma}d\gamma d\phi}{\int^{\phi_\mathrm{max}}_{0}\int^{\gamma_\mathrm{max}}_{0}\cos{\gamma}\sin{\gamma}d\gamma d\phi},
	\label{eps_theta_phi}
\end{equation}
where $\epsilon_{\tilde{\nu}, \gamma}(\tilde{\nu}, \gamma, \phi, T)$ is the spectral directional emissivity. Assuming the emissivity is independent of azimuthal angle $\phi$, eq. \ref{eps_theta_phi} simplifies to:
\begin{equation} 
	\epsilon_{\tilde{\nu}}(\tilde{\nu},T)=\frac{\int_0^{\gamma_\mathrm{max}}\epsilon(\tilde{\nu}, \gamma, T)\cos\gamma\sin\gamma d\gamma}{\int^{\gamma_\mathrm{max}}_{0}\cos{\gamma}\sin{\gamma}d\gamma},
	\label{eps_theta}
\end{equation}
where $\epsilon_{\tilde{\nu}, \gamma}(\tilde{\nu}, \gamma, T)$ is the spectral directional emissivity of the multilayer stack, and $\gamma_\mathrm{max}$ is the maximum angle considered in the integration. 

\begin{figure}
	\includegraphics{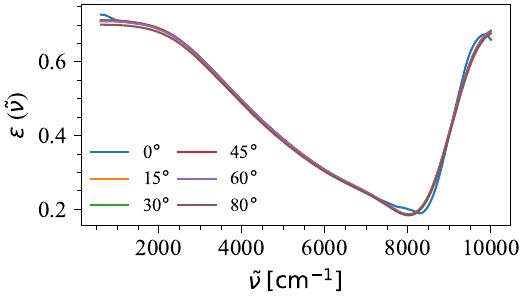}
	\caption{\label{fig:Si_angle} Emissivity of the silicon radiator integrated over various cone angles. 0$^{\circ}$ indicates normal incidence emissivity. Note that this calculation suffers from fast oscillation due to optical interference between the silicon interfaces. Those are smoothed out by a low pass filter.}
\end{figure}

Fig.~\ref{fig:Si_angle} shows the integrated emissivity at room temperature for various $\gamma_\mathrm{max}$. The weak dependence of Fig.~\ref{fig:Si_angle} on $\gamma_\mathrm{max}$ indicates that using the diffused surface emissivity approximation is appropriate. 

\begin{figure}
	\includegraphics{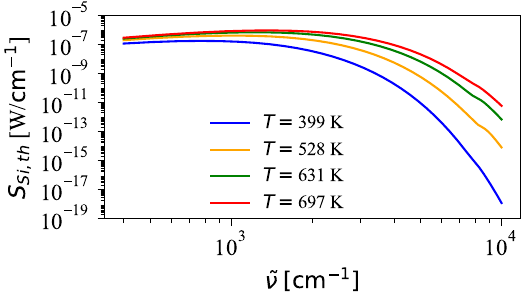}
	\caption{\label{fig:Si_Th} Example theoretical emission of the thermal emitter $S_\mathrm{Si,th}$ with the throughput $\theta=7.24\times10^{-7}$ $\mathrm{m^2\cdot sr}$ for an aperture of $D_\mathrm{ap}=4$ mm and at the temperature used for the MCT detector in CW.}
\end{figure}

With our known emissivity, we can apply eq \ref{S_th} to calculate the theoretical spectral emissivity $S_{\mathrm{th}}(\tilde{\nu})$. Fig.~\ref{fig:Si_Th} shows $S_\mathrm{Si,th}$ for the MCT detector in CW, at emiter temperature taken from Fig.~\ref{fig:MCT}a, and with the throughput $\theta=7.24\times10^{-7}$ $\mathrm{m^2\cdot sr}$ for an aperture of $D_\mathrm{ap}=4$ mm that is used for the MCT detector in CW.

\setcounter{section}{2} 
\renewcommand{\thesection}{\Alph{section}}

\section{LED Emission Test Measurements}\label{appendix_C}

\renewcommand{\thefigure}{\thesection\arabic{figure}}
\renewcommand{\theequation}{\thesection\arabic{equation}}
\setcounter{figure}{0}
\setcounter{equation}{0}

\begin{figure}
	\includegraphics{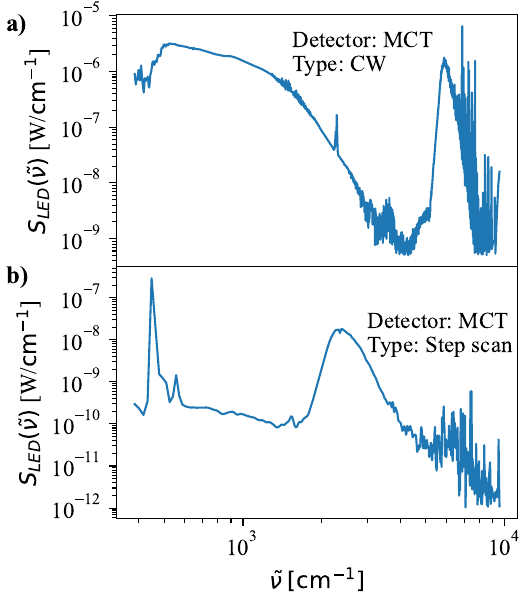}
	\caption{\label{fig:LED_MCT} a) 6060 cm\textsuperscript{-1} LED measurement in CW with MCT detector, where the LED emits 1.07 mW from 5500 cm\textsuperscript{-1} to 7100 cm\textsuperscript{-1}. b) 2325 cm\textsuperscript{-1} LED measurement in step-scan with MCT detector, where the LED emits 0.011 mW from 1850 cm\textsuperscript{-1} to 3800 cm\textsuperscript{-1}. The aperture is fully open ($D_\mathrm{ap}=8$ mm) in both cases.}
\end{figure}

\begin{figure}[H]
	\includegraphics{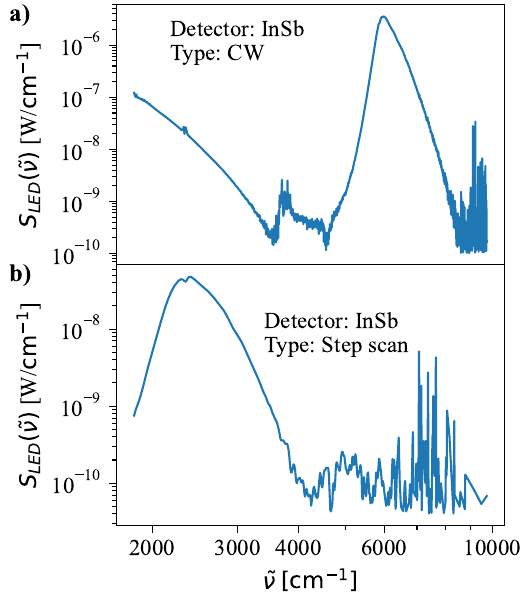}
	\caption{\label{fig:LED_InSb} a) 6060 cm\textsuperscript{-1} LED measurement in CW with InSb detector, where the LED emits 2.19 mW from 5500 cm\textsuperscript{-1} to 7100 cm\textsuperscript{-1}. b) 2325 cm\textsuperscript{-1} LED measurement in step-scan with InSb detector, where the LED emits 0.029 mW from 1850 cm\textsuperscript{-1} to 3800 cm\textsuperscript{-1}. The aperture is fully open ($D_\mathrm{ap}=8$ mm) in both cases.}
\end{figure}

Fig.~\ref{fig:LED_MCT} - \ref{fig:LED_InSb}a presents the emission of a 6060 cm\textsuperscript{-1} LED $S_\mathrm{LED}$ in CW for the MCT and InSb detectors respectively. For the MCT, we measure an emission of 1.07 mW and for the InSb we measure 2.19 mW (from 5500 cm\textsuperscript{-1} to 7100 cm\textsuperscript{-1}), which are within the nominal range of 5 mW. Fig.~\ref{fig:LED_MCT} - \ref{fig:LED_InSb}b presents the emission of a 2325 cm\textsuperscript{-1} LED of 0.8 mW, in step-scan for the MCT and InSb detectors respectively. For step-scan measurement the LED current is modulated between 0 - 100 mA (0 - 0.8 mW) at 1 kHz. With the MCT, we measure an emission of 0.011 mW and for the InSb we measure 0.027 mW (from 1850 cm\textsuperscript{-1} to 3800 cm\textsuperscript{-1}). As expected, this is lower than the nominal value since this particular LED emits most of its radiation within a 30$^{\circ}$ cone that is wider than acceptance angle of the objective. 

\end{document}